# Analogue Forecast System for Daily Precipitation Prediction Using Autoencoder Feature Extraction: Application in Hong Kong

Yee Chun TSOI[1], Yu Ting KWOK[2], Ming Chun LAM[2], Wai Kin WONG[2]

[1]The Hong Kong University of Science and Technology, Clear Water Bay, Kowloon, Hong Kong, China  
[2]Hong Kong Observatory, 134A Nathan Road, Tsim Sha Tsui, Kowloon, Hong Kong, China

*Corresponding author email: ytkwok@hko.gov.hk

ABSTRACT: In the Hong Kong Observatory, the Analogue Forecast System (AFS) for precipitation has been providing useful reference in predicting possible daily rainfall scenarios for the next 9 days, by identifying historical cases with similar weather patterns to the latest outputs from the deterministic model of the European Centre for Medium-Range Weather Forecasts (ECMWF). Recent advances in machine learning allow more sophisticated models to be trained using historical data and the patterns of high-impact weather events to be represented more effectively. As such, an enhanced AFS has been developed using the deep learning technique autoencoder. The datasets of the fifth generation of the ECMWF Reanalysis (ERA5) are utilised where more meteorological elements in higher horizontal, vertical and temporal resolutions are available as compared to the previous ECMWF reanalysis products used in the existing AFS. The enhanced AFS features four major steps in generating the daily rain class forecasts: (1) preprocessing of gridded ERA5 and ECMWF model forecast, (2) feature extraction by the pre-trained autoencoder, (3) application of optimised feature weightings based on historical cases, and (4) calculation of the final rain class from a weighted ensemble of top analogues. The enhanced AFS demonstrates a consistent and superior performance over the existing AFS, especially in capturing heavy rain cases, during the verification period from 2019 to 2022. This paper presents the detailed formulation of the enhanced AFS and discusses its advantages and limitations in supporting precipitation forecasting in Hong Kong.

## 1. INTRODUCTION

Hong Kong is a subtropical city located in the coastal areas of southern China (near 22°N and 114°E) with a hot and humid summer from May to September each year. Due to the southwest monsoon and tropical cyclones, thunderstorms and heavy precipitation events often occur during the period. Taking June as an example, the climatological monthly mean rainfall (1991-2020) is 491.5 mm, and the maximum record was 1346.1 mm in 2008. Moreover, the annual mean number of days with heavy rainfall, defined by daily rainfall of 25 mm or above, of the same period is about 30. Therefore, the forecast of significant or heavy rainfall is important in support of the forecasting and warning services of the Hong Kong Observatory (HKO) for the members of the public and stakeholders.

Recent years see the advances of the global numerical weather prediction (NWP) in terms of enhancements in model configurations including horizontal resolution, data assimilation and model physical processes, etc. In complement to NWP, the analogue forecasting technique has been adopted for predicting more detailed and realistic local weather primarily driven by mesoscale dynamics or subgrid scale processes which may not be captured by global NWP models. The word analogue here refers to two states of atmospheric conditions that are relatively similar (Lorenz, 1969). An analogue forecast system (AFS) aims to search for historical weather patterns that are analogous to the latest NWP model outputs, which are assumed to be sufficiently representative of the expected synoptic conditions. With the notion that "history repeats itself", the occurrences and intensities of weather events like thunderstorms and tropical cyclones are supposed to match with historical cases bearing similar weather patterns, a.k.a. analogues (Bergen & Harnack, 1982). These analogues thus can serve as useful references for the expected local weather, range of possible scenarios, and extremity of weather events for the forecasted days.

Analogue forecasting is widely used in operational predictions of precipitation (Hamill & Whitaker, 2006; Ben Daoud et al., 2016; Zhou & Zhai, 2016), temperature (Bergen & Harnack, 1982; Kruizinga & H. Murphy, 1983), extreme events (Chattopadhyay et al., 2020), and short-term climate (Van Den Dool, 1994; Dayon et al., 2015). Furthermore, the method is applied for supporting impact-based forecasts of flooding (Marty et al., 2012; Bellier et al., 2016) and streamflow (P. Salathé, 2003; Caillouet et al., 2022). In most cases, the analogy of large-scale meteorological patterns can be associated with the corresponding observed local meteorological and/or hydrological element.

The first AFS in the HKO was developed in the early 1990s (Poon & Ma, 1992) for weather forecasts. The meteorological variables for the inputs of the predictor included the surface pressure, winds, and temperature at 850 hPa, the geopotential height (Z) at 500 hPa, and the winds at 200 hPa, each with a 2.5° × 2.5° resolution. Similarity metrics expressed as the anomaly correlation coefficient and S1 score were used for selecting analogues from the historical analysis archive.

The AFS was reformulated in 2014 by Chan et al. and has been used for providing daily precipitation forecasting in support of operation at HKO. The forecasted and historical weather patterns are compared in the following three aspects: the geopotential height pattern, gradient of geopotential height, and moisture elements (Chan et al., 2014). The system employs the self-defined similarity scores of the above predictors, aiming to capture precipitation events, especially those with heavy rainfall. Optimised scores determined by the cuckoo search (Yang and Deb, 2009) are then applied to devise the selection criteria of analogues. Although the AFS has shown skills in forecasting daily precipitation intensity and usefulness in capturing various scenarios, its predictions could vary quite substantially or lack continuity with the inputs of model forecast (ECMWF deterministic model run) from different initial times. Additionally, the performance of this AFS in forecasting extreme weather events could have rooms for improvement. Two possible causes of the above shortcomings are due to the underrepresentation of weather patterns by the self-defined variables and the need for more cases for verification during optimisation.

With a longer record of gridded reanalysis data from the fifth generation of the European Centre for Medium-Range Weather Forecasts (ECMWF) Reanalysis (ERA5) and an increasing number of model variables, an enhanced AFS has been developed in this study to overcome the shortcomings of the existing AFS. Moreover, improvements in computer hardware shorten the time for model training and operation, enabling the application of more sophisticated techniques based on machine learning (ML) and optimisation algorithms. The development of an enhanced ML-based AFS for precipitation has therefore become feasible.

ML-based models are becoming increasingly popular in the field of weather forecasting. Unlike the traditional physics-based numerical models, these models are more robust to perturbations and may adapt better to global climate change by generalising hidden patterns in historical data (Holmstrom et al., 2016; Singh et al., 2019). Applications of ML range from the post-processing of observations (Lo and Lau, 2021) and NWP model outputs (Cho et al., 2022; Schulz and Lerch, 2022), nowcasting (Prudden et al., 2020), to purely data-driven global weather models with neural networks (Pathak et al., 2022; Bi et al., 2023; Lam et al., 2023). Amongst ML methods, the autoencoder emerges as a deep-learning-based feature extraction method that is particularly suitable for generating abstract representations for multi-dimensional data and anomaly detections (Gogna & Majumdar, 2018; Meng et al., 2017; Xing et al., 2016). As such, it is believed that features can be learnt from the gridded data predicted by the ECMWF models using autoencoder. In light of the above, this paper attempts to construct an enhanced AFS for precipitation by comparing the feature representations extracted by an autoencoder and applying the optimised scores for the selection of analogues. In the subsequent sections, the training data and the forecast workflow are first introduced, followed by the model verification and performance evaluation, as well as a brief discussion on potential model enhancements in the conclusion.

## 2. DATA AND METHODOLOGY
### 2.1 Data

Three sets of data are used for the enhanced AFS, namely, (a) the gridded reanalysis data (ERA5), (b) the gridded forecast data from the deterministic model of the ECMWF (HRES), and (c) the daily rainfall observation. The data formats, time periods, and their corresponding purpose at various stages of the model training and operation are summarised in Table 1.

### 2.1.1 Reanalysis Data

ERA5 is the fifth-generation global atmospheric reanalysis dataset provided by the ECMWF. The subset of ERA5 data used in this study contains ten parameters: divergence, geopotential height, potential vorticity, relative humidity, relative vorticity, specific humidity, temperature, zonal component of wind (U), meridional component of wind (V), and vertical wind (W). 2-D gridded data of these parameters are obtained for six isobaric levels, namely 1000, 925, 850, 700, 500, and 200 hPa, forming a total of 60 model features. The gridded data extends over an area defined by 0° – 45°N and 90° – 155°E, covering the region of East Asia (Figure 1) where synoptic-scale weather systems affecting Hong Kong (HK) can be sufficiently captured. The reanalysis dataset for the AFS is formed by the ERA5 data of 00 UTC every day from 1979 to 2020. It is used in the training of the autoencoder model and serves as the historical data archive of analogues.

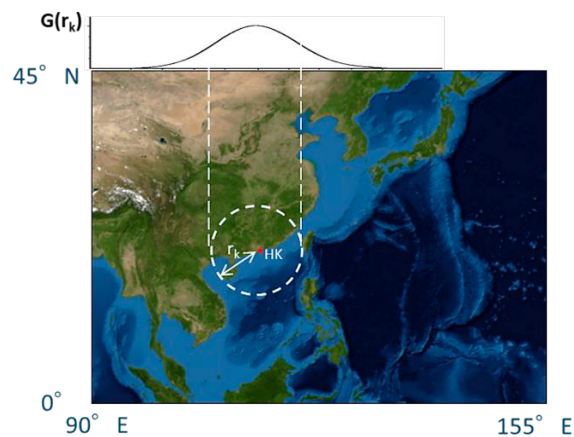

Figure 1: Spatial domain of the gridded data. Also shown is the schematic representation of the Gaussian weight function $G(r_k)$ applied, where $r_k$ is the distance (in degrees) from Hong Kong (HK).

### 2.1.2 Forecast Data

The parameters and spatial information of the ECMWF deterministic model (HRES) data used in this study are identical to that of the reanalysis data. The dataset is formed by forecast data with a model base time of 12 UTC every day, and the forecast valid time of interest is from the Day 1 (T+36 hours) to the Day 9 (T+228 hours) where T denotes the base time of HRES forecast. Day 1 of the forecast data from 2017 to 2020 are used for feature optimisation, while three years of data from May 2019 to April 2022 are used for subsequent forecast verification. For daily operational forecasts, the latest ECMWF HRES model outputs are input into the trained AFS model to be compared with the historical reanalysis data.

Table 1: Overview of the datasets used for model training and operation of the enhanced AFS.

| Dataset | Data Format | Time Period | Purpose (see Section 2.2 for details) |
|---|---|---|---|
| Reanalysis (ERA5) | Gridded numerical (.grb); 0.25° x 0.25° spatial resolution | 1979 – 2020 | • Autoencoder model training<br>• Historical data archive |
| Forecast (ECMWF HRES) | Gridded numerical (.grb); 0.125° x 0.125° spatial resolution | Since 2017 | • Feature optimisation (2017 – 2020)<br>• Verification (May 2019 – Apr 2022)<br>• Operational forecast |
| Daily Rainfall Observation | Numerical values (.csv) | Since 1979 | • Feature optimisation (2017 – 2020)<br>• Verification (May 2019 – Apr 2022)<br>• Operational forecast (rain class calculation) |

### 2.1.3 Rainfall Observation

Another set of data essential for the model training and prediction is the daily rainfall amount in Hong Kong since 1979. The daily rainfall amount is obtained as an average of seven selected reference rain gauges located in different regions in Hong Kong to represent the overall rainfall intensity of the territory. It is used for feature optimisation, forecast verification, and the rain class calculation step of the operational forecast.

## 2.2 Workflow

The model training consists of three main stages: data preprocessing, autoencoder model training, and feature optimisation. On the other hand, the workflow of the daily operational forecast is similar, which includes data preprocessing, feature extraction using the pretrained autoencoder, application of the optimised feature weightings, and a final step to obtain the rain class prediction. The overview of the workflow is illustrated in Figure 2 and further explained below.

### 2.2.1 Data Preprocessing
*2.2.1.1 Reduction of Resolution*

The ERA5 and HRES data have initial spatial resolutions of 0.25° × 0.25° and 0.125° × 0.125° in regular latitude / longitude grid respectively. Before inputting the datasets for model training and operational forecast, the spatial resolution of both datasets is remapped to 1° × 1° by average pooling to harmonize the data resolutions and reduce computational costs of training and inference of the ML model.

*2.2.1.2 Normalisation*

Next, the input gridded datasets are normalised. For each of the 60 model features, values are remapped to a range between 0 and 1 for convenient processing. This would not cause any significant loss of information since the AFS is more concerned with the pattern and relationship between grids rather than the absolute values of each feature. The maximum value of a particular feature within the data archive is set to be 1, while the minimum value is set to 0. The same normalisation process is applied to both gridded datasets in a consistent manner.

*2.2.1.3 Gaussian Weight Function*

With reference to Chan et al. (2014), a Gaussian weight function is applied to assign more weight to grid points closer to Hong Kong and less weight to those further away (Figure 1). In so doing, the AFS can be more sensitive to proximal weather patterns which have the largest effect on local precipitation events. The maximum value of the function is 1 centered at the location of Hong Kong, while its minimum value tends to 0. The equation of the function is as follows:

$$G(r_k) = e^{-\frac{r_k^2}{(1.2 \times r_0)^2}} \quad (1)$$

The term $r_k$ in Equation(1) is the distance of the grid point from Hong Kong in degrees, where 1° equals to about 103 km close to Hong Kong. Furthermore, the term $r_0$ is empirically set to be 7, such that when $r_k$ is equal to $r_0$, the corresponding weight of the grid point $G(r_k)$ will be approximately halved.

### 2.2.2 Autoencoder Model Training and Application

The preprocessed ERA5 data from 1979 to 2020 are shuffled randomly and split into two datasets, with 80% data for training data and 20% data for validation. This strategy ensures that the two datasets can adequately represent the overall distribution of the weather characteristics over the 42 years, so that the exposure of the trained model can be maximised to the historical weather data variability and hence be more robust when handling unseen data. They are used for training the autoencoder, a deep-learning algorithm suitable for extracting meaningful features and detecting anomalies from multi-dimensional data by representation learning (Bank et al., 2020).

An autoencoder consists of an encoder and a decoder, each comprising two convolutional layers and with the inception module applied so to allow feature extraction at multiple scales (Figures 3a and 4). The encoder compresses the 2-D gridded data into 1-D extracted vectors while the decoder attempts to recreate the input with the extracted vectors as much as possible (Bank et al., 2020). The neural network architecture is defined arbitrarily by testing different combinations and minimising the mean squared errors (MSE) between the input and reconstructed images. The training process is automatically stopped when there is no improvement for five consecutive epochs.

The pre-trained autoencoder can generate vectors that represent the feature maps well as illustrated using the example input and output images for the case of 00 UTC 16-09-2018 when severe typhoon Mangkhut skirted around 100 km SSW of Hong Kong (Hong Kong Observatory, 2018). It can be seen that the patterns and magnitudes shown on the original input images are preserved on the corresponding reconstructed image for four selected features (Figure 5). The more alike the input and recreated images are, the more representable the extracted vectors are. Quantitatively, the similarity between two images is represented in terms of MSE.

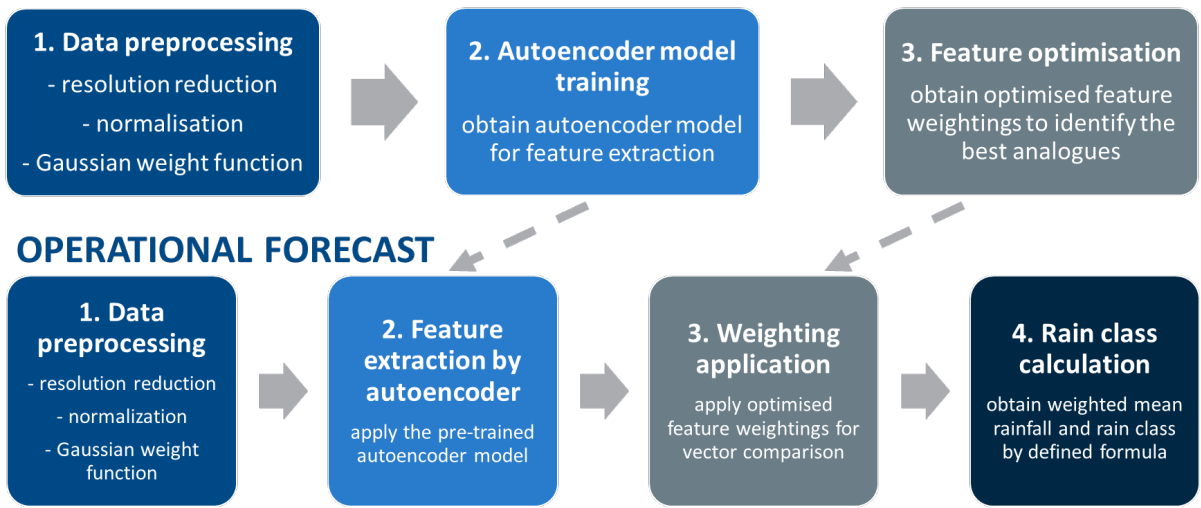

Figure 2: Overview of the AFS workflow for model training and operational forecast.

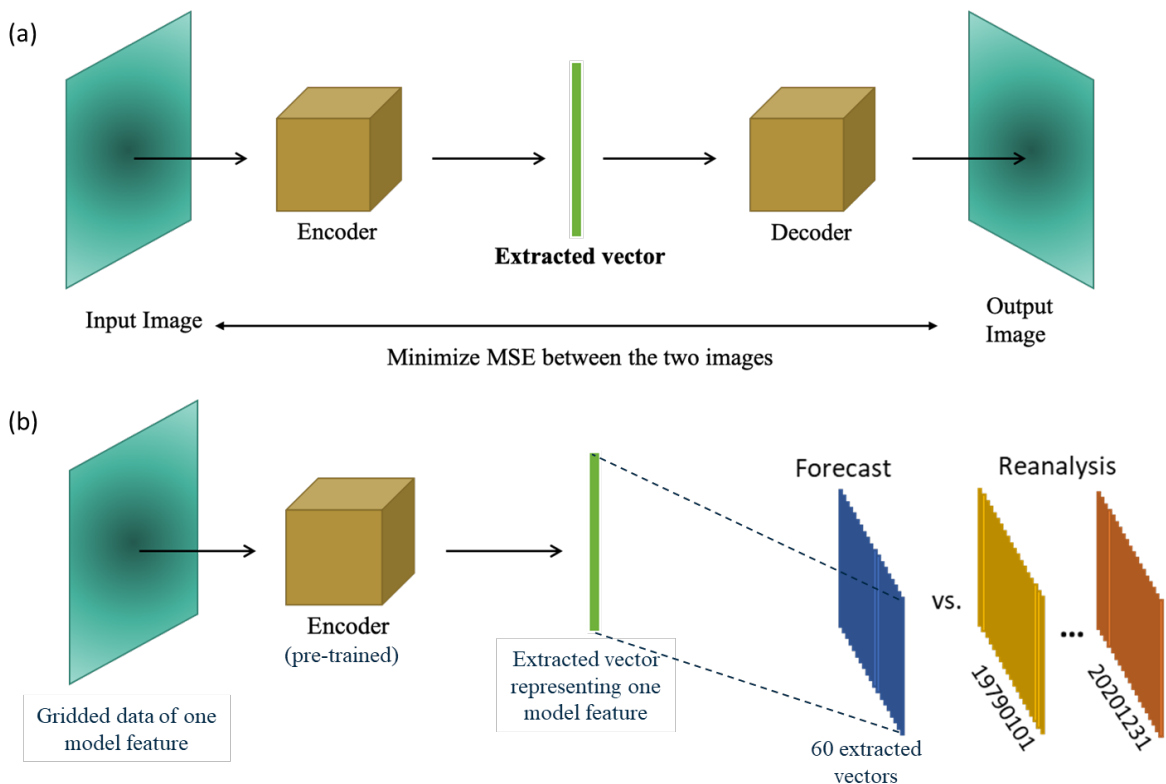

Figure 3: (a) Structure of an autoencoder; (b) Application of the pre-trained autoencoder for operational forecast.

After the autoencoder model is trained, a pre-trained model is obtained for the feature extraction of both ERA5 data archive and HRES forecast data. The extracted vectors of the former are saved in binary format after passing through the autoencoder once to be compared with the extracted vectors from the latest forecast data daily for the operational forecast (Figure 3b). It is noted that the comparison process is restricted to cases within 2 months from the forecasting month. For example, the extracted vectors from the forecast gridded data on any date in December are only compared to cases within the 5-month interval from October to February in the data archive. This is done to ensure that seasonal characteristics of precipitation events are considered when searching for past cases with similar meteorological patterns.

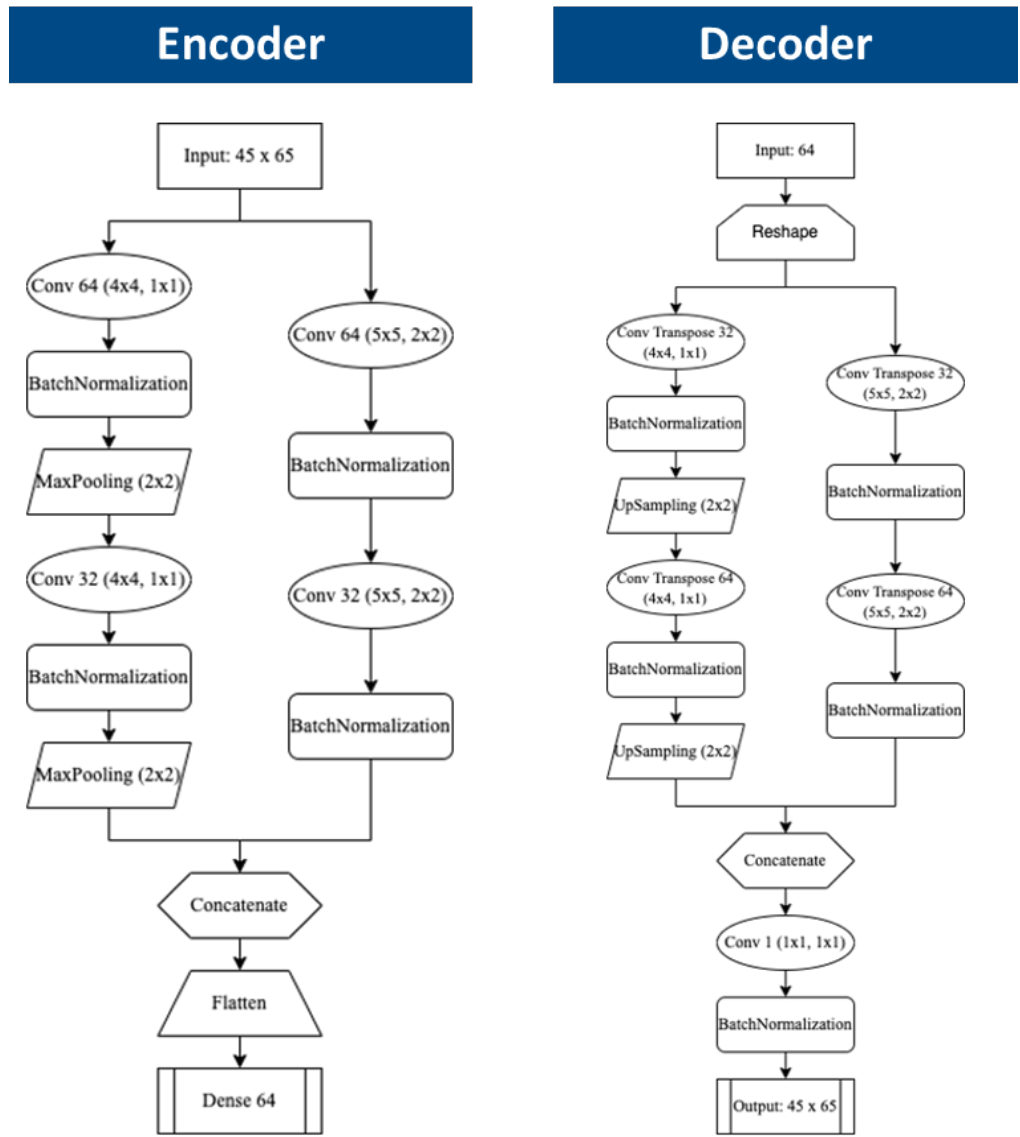

Figure 4: Neural network architecture of the encoder (left) and decoder (right) used in the AFS.

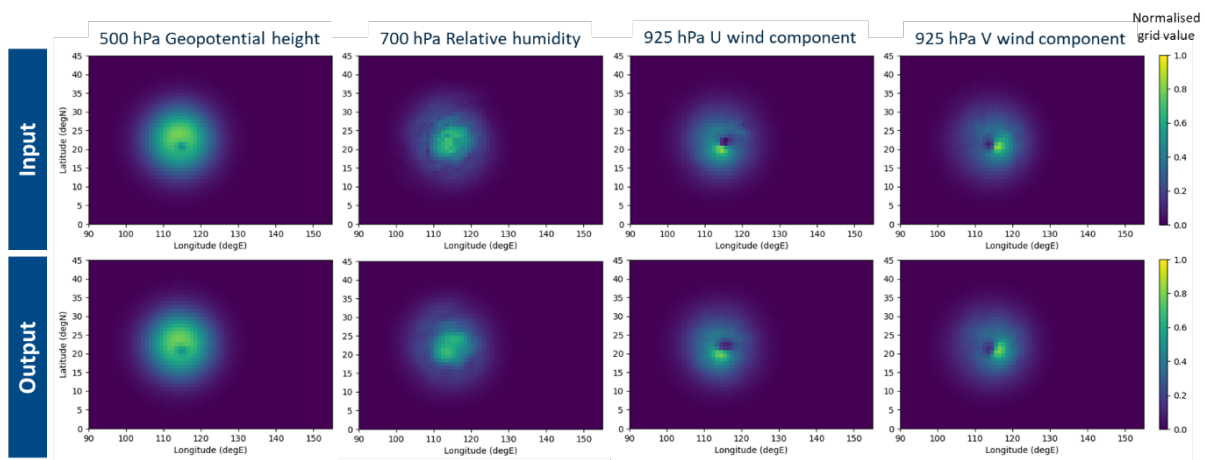

Figure 5: Original and reconstructed images from the pre-trained autoencoder of 500 hPa geopotential height, 700 hPa relative humidity, 925 hPa U wind component, 925 hPa V wind component for the case of 00 UTC 16 September 2018.

*2.2.3 Feature Optimisation and Weightings*

Among the 60 model features, some features may be more relevant and influential to the occurrence and intensity of precipitation events than others. Therefore, it is necessary to deduce an optimised weighting for each feature when identifying the best analogue. The process is performed with a hyperparameter optimisation framework named Optuna (Akiba et al., 2019).

With the pre-trained autoencoder model, the gridded data of the 60 features can be represented with the 60 corresponding extracted vectors every day. The extracted vectors from each day during the optimisation period of 2017 – 2020 are compared to the ERA5 data archive with the same vector expression. MSE are calculated day-by-day and feature-by-feature, and then saved as input files for the optimisation. This comparison process is restricted from 1 January 1979 to 9 days before the verified date and to cases within 2 months from the forecasting month. The normalisation, which remaps MSE values into a range of 0 – 1, is applied to each feature in every file in advance.

During the optimisation, the search space for each parameter is defined at the beginning of each run, with an initial range of 0 – 1 and a step size of 0.001. The objective is to minimise the MSE between the rain classes (Table 2) of each day during the optimisation period and its corresponding best analogue, which is selected from the data archive as the case with least weighted MSE deduced by the feature weightings from the current run.

Each run contains 5000 trials, with the first 1500 trials searching randomly within the space given and the last 3500 trials applying an algorithm of the Tree-structured Parzen Estimator Approach (Bergstra et al., 2011). As the number of trials increases, the parameters start converging while the loss decreases at the same time. At the end of each run, the parameters from the best trial will be obtained and taken as the result of the optimisation from that particular run.

After the first few runs, the range of each parameter will be modified and narrowed down gradually. This step will be performed manually based on the results from previous runs. The process continues until the parameters converge between runs, meaning that the optimised values of the parameters are very similar (within a range of 0.05) for consecutive runs. The resultant optimised weightings of all model features are given in Table 3. These optimised weightings are applied to calculate the weighted MSE and similarity (i.e. 1 / weighted MSE) between each forecast and reanalysis pair during operational forecasts.

Table 2: Definition of rain class by daily rainfall in this study. Classes Heavy, Very Heavy, Torrential, and Extreme are grouped together for model optimisation and verification due to the very few number of cases.

| Rain Class | Daily Rainfall (mm) |
|---|---|
| No Rain | < 0.05 |
| Light | ≥ 0.05 and < 10 |
| Moderate | ≥ 10 and < 25 |
| Heavy | ≥ 25 and < 50 |
| Very Heavy | ≥ 50 and < 100 |
| Torrential | ≥ 100 and < 200 |
| Extreme | ≥ 200 |

*2.2.4 Rain Class Calculation for Operational Forecast*

With the procedures mentioned above, the analogue forecast model can be used to predict daily rain classes for the next 9 days. To avoid large fluctuations of results between model runs and increase the resilience of the model to outliers, an ensemble of similar analogues formed by the 25 most similar analogues is considered in the final rain class calculation. A weighted score (WS) is assigned to each of the 25 analogues according to its weighted MSE and the sample-to-population ratio for each rain class. If the proportion of a certain rain class within the 25 analogues (i.e. sample ratio) is smaller than its occurrence frequency in the past ~20 years (i.e. population ratio; Table 4), the rain class is disregarded from the prediction. The equation, obtained empirically, to calculate the WS (*n* denotes *n*-th analogue) is as follows:

$$WS_n = \begin{cases} \frac{1}{Weighted\ MSE_n} \times \left(\frac{Sample\ Ratio_n}{Population\ Ratio_n}\right)^2, & \frac{Sample\ Ratio_n}{Population\ Ratio_n} \geq 1 \\ 0, & otherwise \end{cases} \quad (2)$$

The weighted scores for the 25 analogues are applied to the corresponding daily rainfall amount (RF) of the analogue to yield the weighted mean rainfall (WMR) for a particular forecast day (Equation (3)). Finally, the predicted rain class is obtained based on the WMR and according to the rain class definitions (Table 2).

$$WMR = \frac{\sum_{n=1}^{25} WS_n \times RF_n}{\sum_{n=1}^{25} WS_n} \quad (3)$$

Table 3: Optimised weightings of all model features for the AFS. Features with the top ten weightings are highlighted in bold.

| Feature \ Layer | 200 hPa | 500 hPa | 700 hPa | 850 hPa | 925 hPa | 1000 hPa |
|---|---|---|---|---|---|---|
| Divergence | 0.011 | 0.106 | 0.560 | 0.568 | **0.941** | 0.666 |
| Geopotential Height | 0.662 | 0.662 | 0.662 | 0.67 | 0.364 | 0.001 |
| Potential Vorticity | 0.095 | 0.004 | 0.09 | 0.407 | 0.017 | 0.246 |
| Relative Humidity | 0.014 | 0.163 | 0.014 | **0.987** | **0.968** | 0.770 |
| Relative Vorticity | 0.322 | 0.053 | 0.057 | 0.736 | 0.807 | **0.968** |
| Specific Humidity | 0.753 | 0.827 | 0.046 | 0.521 | 0.522 | 0.12 |
| Temperature | **0.927** | 0.293 | 0.574 | 0.120 | 0.385 | 0.416 |
| U Wind | 0.67 | **0.933** | **0.905** | **0.968** | 0.656 | 0.647 |
| V Wind | 0.247 | 0.012 | 0.625 | **0.979** | 0.699 | 0.524 |
| Vertical Wind | 0.077 | 0.207 | **0.984** | 0.818 | 0.077 | 0.615 |

Table 4: Statistics on the rain class population ratio based on observations from 2001 to 2020.

| Rain Class | Number of days | Ratio |
|---|---|---|
| No Rain | 1779 | 48.70% |
| Light | 1320 | 36.13% |
| Moderate | 295 | 8.08% |
| Heavy or above | 259 | 7.09% |

## 3. RESULTS AND DISCUSSIONS
### 3.1 Model Verification and Performance Evaluation

The performance of the newly developed ML-based AFS by autoencoder (hereafter, enhanced AFS) is evaluated and compared to that of the AFS (Chan et al., 2014) that has been serving forecasters at HKO for almost a decade (hereafter, existing AFS). The model verification period covers three years from May 2019 to April 2022.

The confusion matrices in Table 5 show the rain class observations against predictions for forecast Day 1, Day 4, and Day 9 from the two AFS. The enhanced AFS is clearly more skillful in capturing heavy rain cases while maintaining similar skill for other rain classes compared to the existing AFS throughout the 9-day forecast period. In particular, over half of the observed heavy rain cases during the verification period are correctly predicted, and this number of cases is more than double of that correctly predicted by the existing AFS. It is worth noting that the enhanced AFS has nil Day 1 forecasts with an error of more than two rain classes, and such unideal cases are also greatly reduced for other forecast days. However, false alarms of 'light rain' when there is 'no rain' increased slightly for the enhanced AFS.

The model performances are also evaluated in terms of verification metrics, namely the critical success index (CSI = hit / (total forecasted + misses)), probability of detection (POD = hits / total observed), and false alarm ratio (FAR = false alarms / total forecasted), for each rain class. Verification metrics are compared between the HKO 9-day Forecast bulletins issued by the forecasters, the enhanced AFS, the existing AFS, and the averaged ECWMF direct model output for grids near Hong Kong in Figures 6-8. With a CSI of almost 0.4 and a POD over 0.5 for forecast Day 1, the enhanced AFS outperforms the forecasters and other models for heavy rain cases from Day 1 to Day 9. Its FAR is also generally lower than the existing AFS. The enhanced AFS also has a similar, if not better, performance compared to the existing AFS in terms of verification metrics for other rain classes. Besides, the CSI for 'no rain' of the enhanced AFS is consistently around 0.2 higher than that of ECMWF direct model output, indicating its ability to correct the NWP model overestimation of rainfall on non-rainy days. Note that the performance of models and forecasters alike appear to be the worst in the predication of 'moderate rain' owing to the narrow range of rainfall amount defined for this rain class (Table 2) and the relatively fewer samples. Nevertheless, for forecast operations considering the impact to the public, it is more important for a model to be able to discern rainy days from non-rainy days, and give forecasters improved guidance for potential days with heavy rain.

Table 5: Confusion matrices of the actual versus forecasted rain classes by the existing and enhanced AFS for forecast Day 1, Day 4, and Day 9. Green cells indicate correct forecasts and grey cells indicate acceptable forecasts of the neighbouring rain class. Refer to Table 2 for definitions of rain classes. Note that the total number of forecasts are slightly different for the two AFS due to missing data.

| Day 1 | Existing AFS (Chan et al., 2014) | | | | Enhanced AFS (this study) | | | |
|---|---|---|---|---|---|---|---|---|
| Fc. / Obs. | No Rain | Light | Moderate | Heavy ≤ | No Rain | Light | Moderate | Heavy ≤ |
| No Rain | 457 | 79 | 6 | 2 | 420 | 112 | 6 | 0 |
| Light | 129 | 205 | 30 | 10 | 75 | 258 | 36 | 13 |
| Moderate | 11 | 34 | 21 | 5 | 4 | 32 | 21 | 16 |
| Heavy ≤ | 7 | 29 | 22 | 16 | 0 | 21 | 14 | 38 |
| Day 4 | | | | | | | | |
| No Rain | 447 | 90 | 3 | 3 | 385 | 130 | 10 | 2 |
| Light | 121 | 209 | 33 | 12 | 71 | 253 | 39 | 16 |
| Moderate | 11 | 36 | 19 | 6 | 6 | 34 | 16 | 19 |
| Heavy ≤ | 6 | 38 | 18 | 12 | 0 | 30 | 21 | 22 |
| Day 9 | | | | | | | | |
| No Rain | 410 | 103 | 14 | 12 | 346 | 136 | 20 | 13 |
| Light | 134 | 182 | 37 | 19 | 94 | 215 | 42 | 28 |
| Moderate | 16 | 37 | 13 | 7 | 6 | 37 | 17 | 12 |
| Heavy ≤ | 13 | 37 | 14 | 10 | 4 | 40 | 11 | 18 |

### 3.2 Heavy Rain Case Study

In view of its promising performance, the enhanced AFS is put into real-time operation and has been launched as an internal forecaster tool (Figure 9) at HKO since May 2022. An episode of heavy rain from 11 to 13 May 2022 is selected to illustrate how the enhanced AFS is able to identify a similar analogue from the historical archive and provide early alert for forecasters regarding the heavy rain event.

A total rainfall of 331.5 mm was recorded during 11 to 13 May 2022, with the rainfall amounts classified as 'very heavy', 'torrential', and 'very heavy' (Table 2), respectively. The enhanced AFS showed clear signals of a 3-day heavy precipitation event up to five days ahead (Figure 9). Looking into more detail for 13 May 2022, the enhanced AFS consistently and accurately predicted 'very heavy' rain since more than a week ahead on 5 May, thus giving forecasters more confidence to include this message in the public forecast bulletins on 7 May.

Meanwhile, the existing AFS gave fluctuating predictions of 'moderate' and 'heavy' for this case in the nine days prior to 13 May.

To further highlight the superior performance of the enhanced AFS compared to the existing AFS, weather charts of selected layers from the corresponding best analogue identified with respect to the ECMWF HRES forecast with model base time at 20220504 12 UTC are shown in Figure 10. The best analogue of the enhanced AFS is a day with 'torrential' rain (11 June 1979). It exhibits more similar synoptic patterns versus the best analogue found by the existing AFS (22 April 2010). Specifically, similarities can be seen in the orientation and location of the surface trough and associated rain band (Figure 10a), the locations of low-level shear line and jets in addition to the moisture content and temperature near Hong Kong (Figure 10b), as well as the narrow ridge and area of strong divergence along the southeastern coast of China at the upper level (Figure 10c).

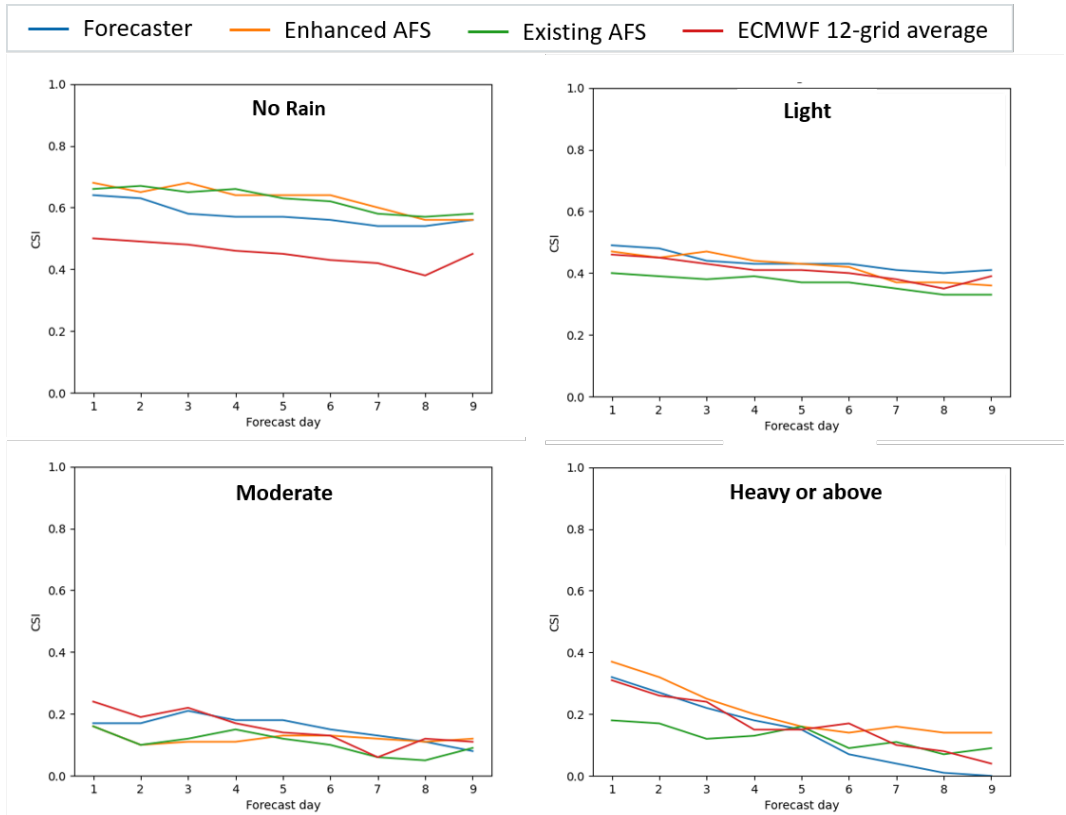

Figure 6: Confusion Comparison of model performance in terms of critical success index (CSI) between HKO forecaster, enhanced AFS, existing AFS, and ECMWF direct model output.

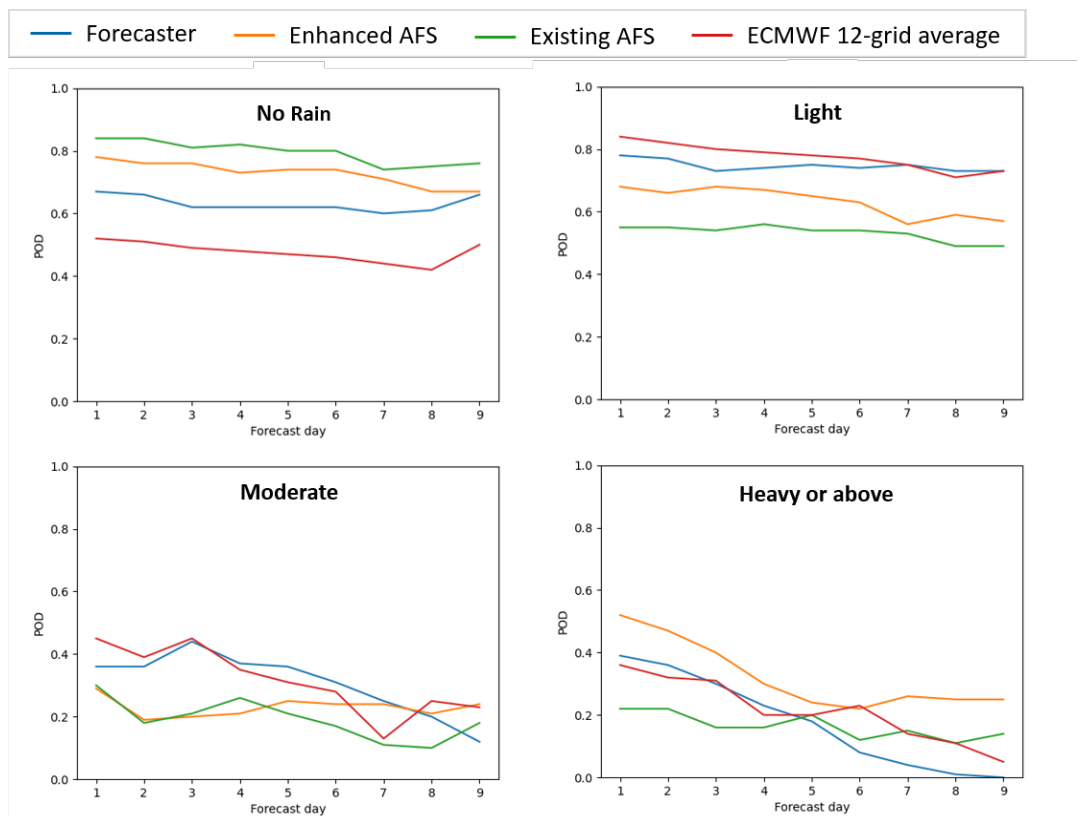

Figure 7: Confusion Comparison of model performance in terms of probability of detection (POD) between HKO forecaster, enhanced AFS, existing AFS, and ECMWF direct model output.

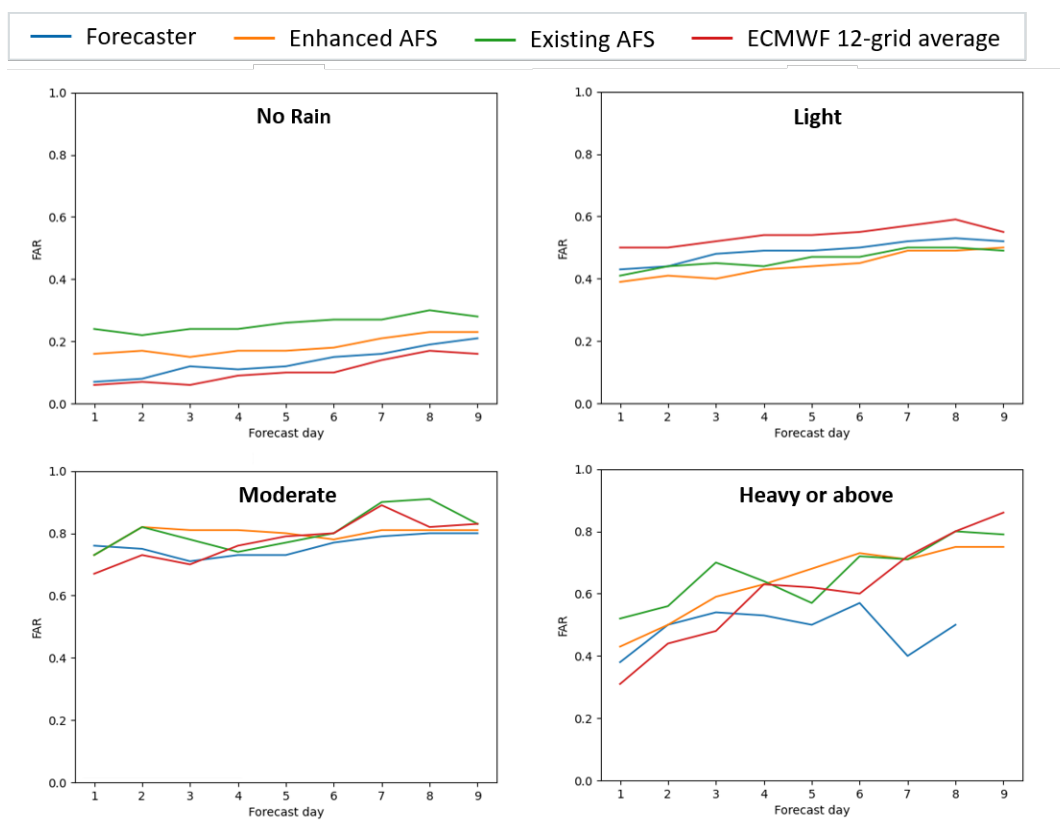

*Figure 8: Confusion Comparison of model performance in terms of false alarm ratio (FAR) between HKO forecaster, enhanced AFS, existing AFS, and ECMWF direct model output.*

### 3.3 Characteristics of the Enhanced AFS

Other than the improvement in performance, major differences between the existing and enhanced AFS are summarised in Table 6. Two features of the enhanced AFS that could benefit the practical application of the model and enhance the understanding of the potential precipitation events are discussed in this section.

### 3.3.1 Meteorological Parameters

Unlike the existing AFS, the enhanced AFS takes advantage of more meteorological parameters included in the HRES and ERA5 data from ECMWF. Parameters such as divergence and relative vorticity, which may indicate favourable conditions for the development of convective systems, are added into consideration in the enhanced AFS. Optimised weightings are also assigned to each parameter so the differential importance of each parameter can be taken into account. This eliminates the insufficient representativeness of the self-defined parameters used in the existing AFS and hence yields more consistent results for the enhanced AFS.

Although the parameter weightings are deduced from a result-oriented optimisation, it can be seen that some highly weighted parameters match with the physical understanding of precipitation processes in a weather forecast. From Table 3, the top ten parameters are temperature on 200 hPa layer, U on 500 hPa layer, U and W on 700 hPa layer, relative humidity, U and V on 850 hPa layer, divergence and relative humidity on 925 hPa layer, and the relative vorticity on 1000 hPa layer. Most of the above results coincide with the self-defined parameters used in the existing AFS and the understanding of the major contributing factors of precipitation, including low-level moisture supply and low- to mid-level synoptic pattern in terms of U and V. Moreover, low-level convergence and relative vorticity relate to convective weather associated with low-pressure systems and tropical cyclones, which often bring heavy rainfall to Hong Kong and the coastal areas of southern China (Wu et al., 2020). These prove that the optimisation results actually show consistency with both the existing AFS and theoretical analyses.

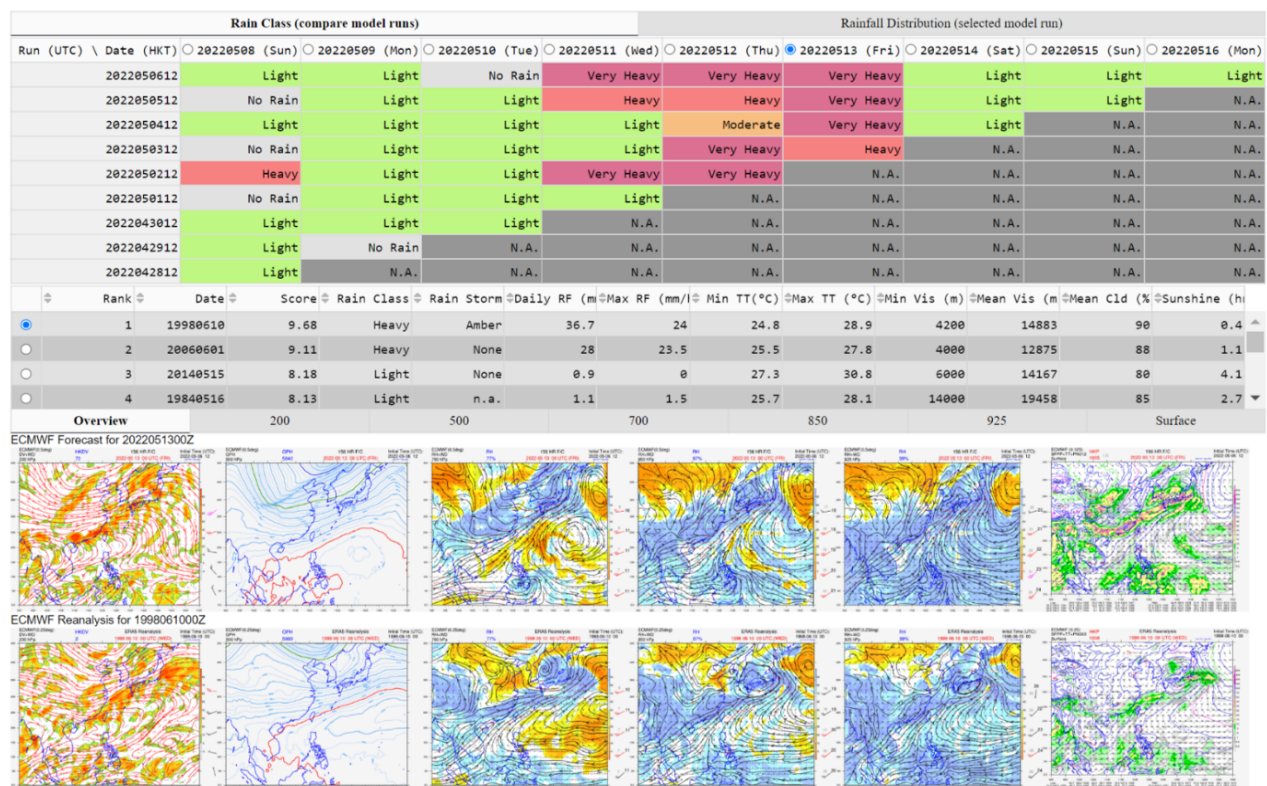

*Figure 9: Web portal of the enhanced AFS internal forecaster tool at HKO. Predictions by the latest and past 9 model runs from 20220506 12 UTC are displayed in the top section; Information of the 25 most similar analogues are listed in the middle section; Forecasted and reanalysed weather charts for the corresponding model run and analogue, respectively, are shown in the bottom section.*

### 3.3.2 Ensemble Forecast

Another modification of the enhanced AFS compared to the existing AFS is the method for obtaining the final rain class output. It considers the top 25 analogues and assigns a score to each analogue for the estimation of the WMR and subsequently the rain class. By considering multiple analogues, the model not only reduces the probability of the output being fluctuated by outliers, but also results in a traceable ensemble forecast with information of the members clearly listed. Furthermore, the rainfall distribution from the 25 analogues is summarised and provided as reference for the forecasters. This could give an indication of the range of anticipated weather scenarios and more insight on the uncertainty of possible weather events when formulating the weather forecast.

### 4. CONCLUSION AND FUTURE WORK

An enhanced AFS has been developed utilising the machine learning framework. There are three main stages in the model training, including data preprocessing, autoencoder model training, and feature optimisation by the Optuna framework. With the aid of machine learning, the implementation of the autoencoder feature extraction is able to capture information from the 60 features of the ECMWF forecasts and find analogues more effectively. For the operational forecast, the pre-trained autoencoder and optimised feature weightings are applied to the latest forecasts. Then, by considering the similarity and past occurrence frequency of the top 25 analogues, a final rain class output is obtained. The enhanced AFS demonstrates promising ability in capturing heavy rain cases and outperforms the existing AFS consistently over the 9-day forecast period during the 3-year verification period. The analogues identified by the enhanced AFS serve as additional information to forecasters on the plausible local rainfall amounts for a similar synoptic pattern depicted by ECMWF HRES, while the final rain class from the system provide guidance on the rainfall intensity summarised from the ensemble of analogues. The new system has been in real-time operation since May 2022 and has

| ECMWF forecast | Best analogue of existing AFS | Best analogue of enhanced AFS |
| --- | --- | --- |
| (20220504 12 UTC T+204h) | (reanalysis of 20100422 00 UTC) | (reanalysis of 19790611 00 UTC) |
| 'Very Heavy' rain class | 'Moderate' rain class | 'Torrential' rain class |

**(a) Surface level**

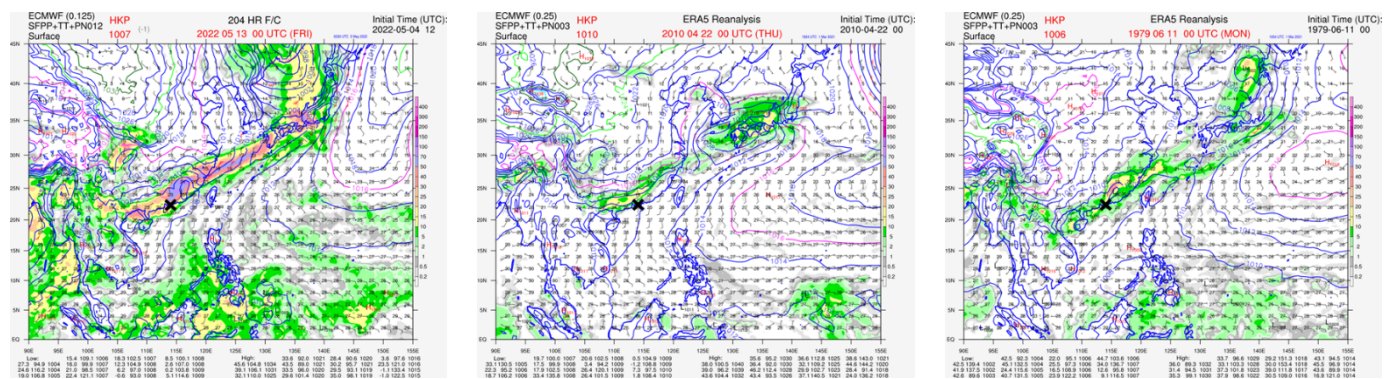

**(b) 925 hPa level**

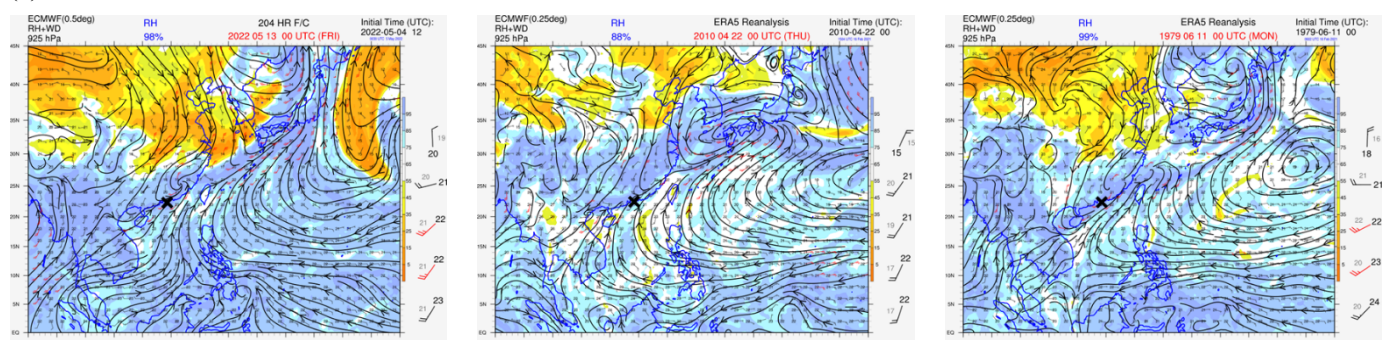

**(c) 200 hPa level**

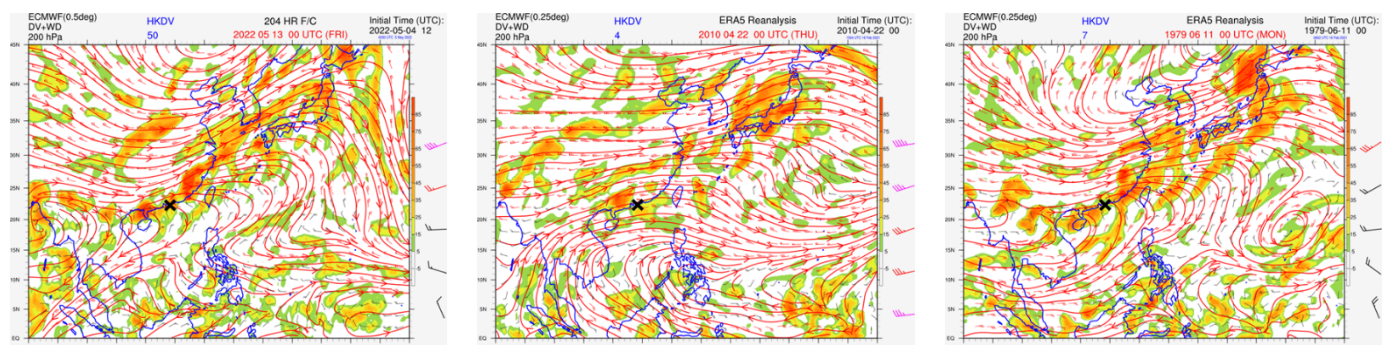

Figure 10: Comparison of (a) surface level, (b) 925 hPa level, and (c) 200 hPa level weather charts from ECMWF forecast with model base time at 20220504 12 UTC against the best analogues found by the existing and enhanced AFS for the case on 13 May 2022. The location of Hong Kong is marked by a black cross.

been well-received by forecasters at HKO, providing useful forecast guidance on the anticipated rain class and early warning for episodes of heavy rain.

Despite the improvements achieved by the enhanced AFS, there are limitations due to its methodology and the complexity of weather systems around the coastal areas of southern China. Events like tropical cyclones, convective storms and cold fronts are very hard to predict in terms of how much precipitation they would be bringing to Hong Kong, as the atmospheric environment may be changing rapidly in a relatively small spatiotemporal scale under these conditions (Chen et al., 2020; Marks et al., 1998). Besides, the performance of an AFS is highly dependent on the skill and accuracy of the NWP model forecasts, in this case the ECMWF HRES data. Another intrinsic weakness of an AFS is its inability to find matching cases for unforeseen extreme weather events brought by climate change. In order to overcome some of these limitations, several modifications could be explored.

*Table 6: Major differences between the existing AFS and enhanced AFS.*

| | Existing AFS (Chan et al., 2014) | Enhanced AFS |
|---|---|---|
| Historical Archive | ERA-Interim reanalysis (1979-2008) | ERA5 reanalysis (1979-2020) |
| Search Window | Forecast date ± 50 days (101 days) | Forecast month ± 2 months (5 months) |
| Model parameters | Geopotential Height, Gradient of Geopotential Height Field (Z at 700, 850 & 925 hPa only), Relative Humidity (at 700 & 850 hPa only) | Divergence, Geopotential Height, Potential & Relative Vorticity, Relative & Specific Humidity, Temperature and U, V & W wind fields |
| Model layers | 200, 500, 700, 850 & 925 hPa | 200, 500, 700, 850, 925 & 1000 hPa |
| Grid Weighting | Gaussian weight function, except for similarity score of Z at 500 hPa (synoptic pattern matching) | Also uses the same Gaussian weight function to give more weight to grid points closer to HK |
| Methodology | Z pattern & gradient matching (anomaly correlation coefficient & S1 score), Moisture matching (comparison of average RH at upstream grids) | Autoencoder feature extraction & vector comparison (in terms of MSE) |
| Optimisation | Cuckoo Search; to determine system parameters that maximize CSI of heavy rain class and minimize cases with no rain class | Optuna; to deduce appropriate weightings for each feature in order to minimize MSE between best analog and actual rain class |
| Rain Class Calculation | From mean rainfall of analogs considered sufficiently similar, otherwise only 1 best analog | From weighted mean rainfall of top 25 analogs weighted based on similarity and rain class population ratio |

### 4.1 Number of Analogues Considered

The number of analogues used to determine the final rain class is set to be 25 based on the tested results. However, there may not be that many cases from the data archive matching particular events that have seldom occurred in the past. This results in a higher chance that some less similar analogs are taken under consideration in this circumstance. Therefore, fewer analogues might be considered for such calculation to prevent biases caused by irrelevant analogues. One potential solution is to vary the number of analogues considered for different seasons. For dry seasons like fall and winter, precipitation is less likely to happen, and hence similar cases may be more challenging to find. Lowering the number of analogues in the ensemble may be preferred. Another possible modification is to apply thresholds to filter less similar analogues. In this case, analogues with weighted mean scores less than the thresholds would not be considered even if they are ranked at the top. However, a more in-depth study would be required to determine the optimal threshold to be adopted for this filtering.

### 4.2 Method to Calculate the Final Rain Class

On some occasions, there may be a wide spectrum of weather scenarios among the top-ranked analogues, which may contribute to the overestimation or underestimation of the WMR. In the WS calculation step, adjustments of parameters like the thresholds of the sample-to-population ratio for each rain class and the relationship between the ratios and the weighted score may be implemented through further optimisations. Nevertheless, this modification would not be effective if the output rain class has been misled by outliers among the top-ranked analogues, probably due to the intrinsic model error from ECMWF HRES forecast data. Besides, care must be taken not to overfit the empirical equation to cases in the verification period.

## 4.3 Increase in Resolution

The data used for the model is currently reduced to a spatial resolution of 1° × 1°. Although the reduction of data size lowers the running time of multiple procedures, there may be loss of information during the process. As Hong Kong is small compared to the grid size of the NWP models, the gridded data may not be reflecting sufficient details for some small-scale atmospheric processes leading to precipitation. Therefore, one way to improve the model is to maximise the resolution to the gridded data while maintaining an acceptable efficiency of the model. Since most parts of the models are optimised and finalised in the model training, it is feasible to maintain high efficiency for the operational forecasts even with finer gridded data, given advancements in computer hardware.